\documentclass[a4paper,11pt]{article}
\usepackage[utf8]{inputenc}
\usepackage{jheppub} 
\UseRawInputEncoding
\usepackage{lineno}
\usepackage{physics}
\RequirePackage{verbatim}
\NewDocumentCommand{\tens}{t_}
{%
	\IfBooleanTF{#1}
	{\tensop}
	{\otimes}%
}
\NewDocumentCommand{\tensop}{m}
{%
	\mathbin{\mathop{\otimes}\displaylimits_{#1}}%
}
\usepackage{amsmath}
\usepackage{color}
\usepackage{amsfonts}
\usepackage{verbatim}
\usepackage[toc,page]{appendix}
\usepackage{amssymb}
\usepackage{bbold}
\usepackage{graphicx}
\usepackage{pgfplots}
\pgfplotsset{compat=1.18}
\usepackage{float}
\usepackage{float}
\title{Can Non-Relativistic Strings Propagate Without Geometric Baggage?\\
\large A Minimal Hamiltonian Formulation in Newton--Cartan Geometry}
\author[1,a,b]{Partha Nandi,\note{Corresponding author}}
\author[c]{Sk. Moinuddin,}
\author[d]{Abdus Sattar}
\affiliation[a]{Department of Physics, \\University 
	of Stellenbosch, Stellenbosch-7600, South Africa}
\affiliation[b]{National Institute for Theoretical and Computational Sciences (NITheCS),\\Stellenbosch, 7604, South Africa}
\affiliation[c]{Department of Physics,\\ Darjeeling Government College, Darjeeling, India}
\affiliation[d]{Tapna Chaturbhuj High School,\\
Balakhali Tapna, Bishnupur, \\South 24 pgs, West Bengal-743503, India}

\emailAdd{pnandi@sun.ac.za} \emailAdd{dantary95@gmail.com} \emailAdd{sat.abdus@gmail.com}

\abstract{ 

We present a minimal and dynamically consistent formulation of non-relativistic bosonic string theory in a  Newton-Cartan  background. Starting from a reparametrization-invariant Nambu-Goto action, we develop the Hamiltonian framework and perform a complete Dirac constraint analysis. The resulting structure exhibits first-class constraints that generate worldsheet diffeomorphisms, confirming the internal gauge consistency of the model. Using an interpolating Lagrangian, we derive a Polyakov-type action that enables a direct comparison with symmetry-based constructions known as gauging-the-algebra (GTA) approaches, which promote non-relativistic symmetry algebras to local symmetries. In contrast to GTA formulations which require additional background fields to achieve algebraic closure, our model derives all necessary geometric data dynamically from the string evolution itself. This establishes that standard  geometry is sufficient to support consistent non-relativistic string dynamics. Our results provide a conceptually transparent and technically robust foundation for future studies of non-relativistic string theory in curved backgrounds.}

\begin{document}
	\maketitle
	\flushbottom

    \section{Introduction}

Recent years have witnessed a notable resurgence of interest in the non-relativistic limits of field and string theories~\cite{Oling:2022fft,Kunzle:1976}, spurred by developments in flat-space holography~\cite{Bousso:2004kp}, the physics of strongly correlated systems~\cite{Fulde:1997ijj}, and the ongoing quest for coherent formulations of non-Lorentzian gravity~\cite{Bergshoeff:2022iyb,FrederikGScholtz:2025mcs}. In this context, a compelling paradigm was recently advanced in~\cite{Nandi:2023tfq}, where a hidden Lorentz covariance was uncovered within the Schr\"odinger equation formulated on three-dimensional Euclidean space. This striking observation invites a deeper conceptual inquiry: is it truly necessary to treat space and time on equal footing in relativistic theories, or might Lorentz symmetry emerge more subtly than traditionally assumed?

A recurring theme across many of these investigations is the use of  Newton-Cartan (NC)  geometry~\cite{Cartan:1923,Cartan:1924}, which provides a covariant reformulation of Newtonian gravity \cite{misner1973gravitation} well-suited for coupling to non-relativistic matter fields~\cite{Mukherjee:2018cka, Fernandes:2017ghz}, including in noncommutative settings~\cite{Muthukumar:2004wj,Bose:2024yfe}. A foundational treatment of NC geometry coupled to matter was developed by Banerjee et al.~\cite{Banerjee:2015rca}, whose work laid the groundwork for further developments. While Banerjee's analysis focused on geometric structure and coupling mechanisms, a distinct application of NC geometry was employed by Son~\cite{Son:2013rqa} in the effective field theory description of the fractional quantum Hall effect. These complementary lines of investigation have since inspired a range of related studies~\cite{Banerjee:2015paa, Banerjee:2014nja, Banerjee:2015twa, Banerjee:2020wbr}.

While the dynamics of point particles in Newton--Cartan (NC) backgrounds are now well understood~\cite{Banerjee:2020bym}, the behavior of extended objects such as strings---which can probe deeper layers of spacetime geometry---remains comparatively less explored, opening important directions for theoretical investigation.

Non-relativistic strings~\cite{Oling:2022fft} offer a more refined probe of background structure than point particles: whereas particles respond primarily to local features along their worldlines (such as lapse and shift), strings sweep out two-dimensional worldsheets that are sensitive to geometric structures like torsion, foliation, and transverse--longitudinal couplings. These features play a central role in non-AdS holography and condensed matter systems with emergent non-relativistic symmetries~\cite{PhysRevD.111.026003}, where geometry often resists reduction to geodesic motion~\cite{Karan:2024jgn}. In this context, several semiclassical studies have examined non-relativistic string configurations in torsional NC backgrounds, highlighting integrability and rigidly rotating solutions~\cite{Roychowdhury:2020nonrel_integrable,Roychowdhury:2020giant_magnons}.

To describe this more intricate interaction between non-relativistic strings and background geometry, various geometric extensions have been proposed specifically for strings. One prominent framework is the so-called \textit{stringy Newton--Cartan} (SNC) geometry~\cite{Andringa:2012uz,Bergshoeff:2014uea,Bergshoeff:2019pij}, which arises from gauging non-relativistic symmetry algebras such as the stringy Galilei or Newton--Hooke algebra~\cite{Pal:2024btm}. These gauging-the-algebra (GTA) constructions are grounded in elegant symmetry principles but typically require auxiliary background fields---such as extension gauge fields $m_\mu^a$ and $m_\mu^{ab}$---to ensure algebraic closure~\cite{Andringa:2012uz}. While symmetry-consistent, the physical necessity and dynamical role of these additional structures remain open to question.

This leads to a fundamental issue: \textit{Can non-relativistic string dynamics be formulated consistently without invoking algebraically imposed geometric extensions?} In this paper, we answer this in the affirmative by constructing a reparametrization-invariant action for a non-relativistic bosonic string propagating in a general NC background and analyzing its constraint structure from first principles. Our formulation avoids extraneous gauge fields and derives all consistency conditions dynamically from the canonical evolution of the string.

\paragraph{Our standpoint.}
The approach we take in this work is dynamically minimal and variational in nature. Unlike GTA-based models-which begin with a non-relativistic algebra and gauge all its generators to build the geometry from symmetry closure-we start from a fixed  background and construct a reparametrization-invariant string action consistent with that geometry. We do not gauge spacetime symmetries in the conventional sense, nor do we impose closure of a symmetry algebra. Instead, the  background provides the minimal geometric data necessary for coupling, and we derive the full constraint structure through Hamiltonian analysis. This reveals dynamical constraints on the background geometry that go beyond symmetry arguments and shows that auxiliary extension fields are not needed for consistency. Our formulation is thus distinct from both algebraic and extended geometric approaches, and offers a conceptually transparent foundation for non-relativistic string theory.

Our approach is further motivated by recent work~\cite{PhysRevD.103.046020, Moinuddin:2021oft}, which questions certain assumptions underlying GTA-based models-such as their curvature constraints and foliation structure-that may obstruct a consistent Hamiltonian treatment. In line with this, we find that the constraint algebra of the non-relativistic string imposes dynamical conditions on the background that cannot be inferred from symmetry closure alone.

Let us also emphasize a key conceptual departure from earlier Hamiltonian analyses such as~\cite{Andringa:2010it}, where the background geometry is treated as externally fixed and non-dynamical, allowing for symmetry classification but not dynamical feedback. In our framework, the background fields are treated as dynamical, and their evolution is subject to the constraints imposed by string motion. This broader setting reveals additional geometric interdependencies and consistency conditions that are invisible in fixed-background approaches.

In the course of our analysis, we demonstrate that:
\begin{itemize}
    \item Canonical consistency leads to geometric constraints beyond those required by symmetry closure;
    \item Several curvature constraints imposed in GTA constructions must be relaxed or reinterpreted to preserve Hamiltonian evolution;
    \item Torsion and auxiliary gauge fields play different roles when geometry is derived from variational dynamics;
    \item A revised notion of SNC geometry may be needed-one rooted in dynamics rather than symmetry algebra.
\end{itemize}

Our results thus provide a minimal, dynamically complete alternative to existing SNC constructions and a physically grounded framework for studying non-relativistic string propagation in curved backgrounds.

The present model builds upon foundational developments in non-relativistic gravity~\cite{Banerjee:2015paa, Banerjee:2014nja, Banerjee:2015twa, Banerjee:2020wbr}, where  geometry is coupled to matter using first-order vielbein-based formalisms. The degeneracy of the NC metric renders second-order (Einstein-like) formulations inadequate and necessitates a careful treatment of torsion, foliation, and constraints. This framework remains fully consistent with that philosophy and extends it to a string-theoretic setting.


A central result of the paper is the derivation of a Polyakov-type action using an interpolating Lagrangian approach, enabling a transparent reformulation in covariant worldsheet language. This allows direct comparison with symmetry-based SNC models and reveals that many of the additional fields introduced in GTA approaches are dynamically redundant. We find that these extensions are not necessary—at least in the classical setting explored here—for consistent string dynamics.

\paragraph{Organization of the paper.}
The remainder of this paper is structured as follows. In Section~2, we develop the Lagrangian formulation of the non-relativistic bosonic string. We begin by reviewing the flat spacetime case in Section~2.1, and then extend the formulation to curved Newton-Cartan backgrounds in Section~2.2. Section~3 is devoted to the Hamiltonian analysis, where we carry out a gauge-independent constraint classification in Section~3.1, followed by a gauge-fixed treatment in Section~3.2. In Section~4, we derive the Polyakov-type action using an interpolating Lagrangian. Section~5 presents a detailed comparison with earlier approaches based on gauged non-relativistic algebras. We conclude in Section~6 with a summary and outlook.

\section{Minimal String Dynamics: The Lagrangian Perspective}

\subsection{Non-Relativistic String in Flat Target Space}

To set the stage for our formulation in curved backgrounds, we begin by reviewing the Lagrangian dynamics of a bosonic string in flat spacetime and examine its non-relativistic limit. The bosonic string is a one-dimensional relativistic object whose motion sweeps out a two-dimensional worldsheet. This worldsheet is parametrized by coordinates \( \sigma^\alpha = (\tau, \sigma) \), where \( \tau \) is the worldsheet time and \( \sigma \) is the spatial parameter along the string.

The standard relativistic dynamics of the bosonic string is governed by the Nambu-Goto action:
\begin{equation}
S_{\text{NG}} = -T \int d\tau\, d\sigma \sqrt{-\det h_{\alpha\beta}},
\end{equation}
where \( T \) is the string tension and \( h_{\alpha\beta} \) is the induced metric on the worldsheet, given by:
\begin{equation}
h_{\alpha\beta} = \eta_{AB} \, \partial_\alpha X^A \, \partial_\beta X^B, \qquad A, B = 0, 1, \dots, D.
\end{equation}

Here, \( X^A(\tau, \sigma) \) are the embedding coordinates of the string in target spacetime, and \( \eta_{AB} = \text{diag}(+1, -1, \dots, -1) \) is the Minkowski metric.

To obtain a non-relativistic string theory, we consider the formal limit \( c \to \infty \), effectively decoupling temporal and spatial directions. This process breaks Lorentz invariance and introduces a natural split between the longitudinal directions \( X^\mu = (X^0, X^1) \) and the transverse directions \( X^a \), with \( a = 2, \dots, D \).

In this limit, the Nambu-Goto action reduces to a non-relativistic Lagrangian of the form:
\begin{equation}
\mathcal{L}_{\text{NR}} = -T \left( \epsilon_{\mu\nu}  \, \partial_\tau X^\mu \, \partial_\sigma X^\nu \right)^{-1} \left( \epsilon^{\alpha\beta} \, \partial_\alpha X^\mu \, \partial_\beta X^a \right)^2.
\label{ng1}
\end{equation}

Here, \( \epsilon^{\alpha\beta} \) is the antisymmetric Levi-Civita symbol on the worldsheet, with \( \epsilon^{\tau\sigma} = 1 \), and \( \mu, \nu \in \{0,1\} \), while \( a = 2, \dots, D \) denotes the transverse directions.

This action describes a non-relativistic string in flat spacetime with Galilean symmetry in the target space. The factor \( \left( \epsilon_{\mu\nu} \, \partial_\tau X^\mu \, \partial_\sigma X^\nu \right)^{-1} \) plays a role similar to the induced metric determinant in the relativistic case, ensuring invariance under worldsheet reparametrizations.

\subsection{Extending to Curved Newton-Cartan Geometry}

To extend the non-relativistic string action to a gravitational background, we consider the low-energy regime in which gravity itself behaves non-relativistically. In this framework, the structure of the worldsheet remains unchanged, since the string remains an inherently relativistic object~\cite{Polchinski1998}. Consequently, the two-dimensional worldsheet describing the evolution of the string through spacetime retains a Minkowski metric, while the transverse bulk spacetime exhibits a Euclidean structure.

Following this approach, the Galilean Gauge Theory (GGT) algorithm \cite{Banerjee:2018gqz, Banerjee:2017jyb} provides a systematic framework for constructing the string action in a curved manifold. The key modification involves replacing ordinary derivatives with covariant derivatives in the transverse directions.

It is important to note that the action for the non-relativistic bosonic string remains invariant under the following global transformations:
\begin{equation}
X^{A} \to X^{A} + \xi^{A},
\label{1008}
\end{equation}
where the transformation parameters are given by:
\begin{align}
\xi^0 &= -\epsilon, \\
\xi^1 &= \epsilon^{1} - v^1 X^0, \\
\xi^a &= \epsilon^a + \omega^a_{\;l} X^l - u^a X^{0}, \quad a = 2, \dots, D.
\end{align}

However, upon promoting the infinitesimal parameters \( \epsilon^\mu \) and \( \omega^\mu_{\;~\nu} \) to local functions, the action is no longer invariant under these transformations. To restore local invariance under Lorentz-like transformations, it becomes necessary to introduce additional compensating fields. This requires replacing ordinary derivatives with covariant derivatives, ensuring that the action remains locally invariant in the transverse sector.

The action for a non-relativistic limit of bosonic string in a curved background then takes the form~\cite{Moinuddin:2021oft}:
\begin{equation}
S_{NG} = -T \int d\sigma d\tau \left( \epsilon_{\mu \nu} \sigma_{\alpha\beta} \frac{\partial X^{\mu}}{\partial\sigma_{\alpha}} \frac{\partial X^{\nu}}{\partial\sigma_{\beta}} \right)^{-1} \left( \epsilon_{\alpha \beta} \frac{\partial X^{\mu}}{\partial\sigma_{\alpha}} \frac{D X^{a}}{d\sigma_{\beta}} \right)^2.
\label{1007}
\end{equation}

where \(\sigma^{\alpha \beta}\) is the matrix with components \(\sigma^{01} = \sigma^{10} = 1\) and zero otherwise.

Here, the covariant derivative is defined as

\begin{align}
\frac{DX^{a}}{d\sigma_{\beta}} &= \frac{\partial X^{l}}{\partial\sigma_{\beta}} \Lambda_{l}{}^{a}, \nonumber \\
\delta \left( \frac{DX^{a}}{d\sigma_{\beta}} \right) &= \frac{D X^{b}}{d\sigma_{\beta}} \omega^a{}_b - \frac{\partial X^{0}}{\partial \sigma_{\beta}} u^{a}, \nonumber \\
\delta \left( \frac{\partial X^{\mu}}{\partial\sigma_{\alpha}} \right) &= \frac{\partial X^{\nu}}{\partial\sigma_{\alpha}} \omega^\mu{}_\nu.
\label{n1}
\end{align}

The transformation law for the newly introduced fields~\cite{Moinuddin:2021oft} is:
\begin{equation}
\delta \Lambda_l{}^a = \omega^a{}_b \Lambda_l{}^b - \partial_l\xi^m \Lambda_m{}^a.
\label{old1x}
\end{equation}
All other fields remain invariant under these transformations.

It is worth mentioning that, in the flat-space limit, the fields \( \Lambda_{l}{}^{a} \) reduce to Kronecker deltas. Substituting this into Eq.~\eqref{1007}, and using the transformation laws in Eq.~\eqref{n1}, the action reduces to the flat-space non-relativistic string action in Eq.~\eqref{ng1}. Thus, our construction is consistent in the flat limit.

In the GGT framework, the fields \( {\Lambda_\rho}^{\gamma} \) can be interpreted as inverse vielbeins in a curved manifold. We define \( \Sigma_\gamma{}^\sigma \) as the inverse of \( {\Lambda_\rho}^{\gamma} \), satisfying:
\begin{equation}
\Sigma_\gamma{}^\rho {\Lambda_\rho}^{\epsilon} = \delta^\epsilon_\gamma, \qquad \Sigma_\gamma{}^\rho {\Lambda_\sigma}^{\gamma} = \delta^\rho_\sigma.
\label{B}
\end{equation}

These vielbeins establish a correspondence between global and local bases:
\begin{equation}
\hat{e}_\rho = \Lambda_\rho{}^\gamma \hat{e}_\gamma, \qquad \hat{e}_\gamma = \Sigma_\gamma{}^\rho \hat{e}_\rho.
\label{basis}
\end{equation}

As shown in~\cite{Moinuddin:2021oft}, defining
\begin{equation}
h^{\rho\sigma} = \Sigma_i{}^{\rho} \Sigma_i{}^{\sigma}, \qquad \tau_{\rho} = \Lambda_\rho{}^{0},
\label{spm}
\end{equation}
and
\begin{equation}
h_{\rho\sigma} = \Lambda_{\rho}{}^{i} \Lambda_{\sigma}{}^{i}, \qquad \tau^{\rho} = \Sigma_0{}^{\rho},
\label{spm2}
\end{equation}
leads to the Newton-Cartan compatibility relations:
\begin{equation}
h^{\rho\sigma} \tau_\sigma = 0, \quad h_{\rho\sigma} \tau^\sigma = 0, \quad \tau^\rho \tau_\rho = 1, \quad h_{\rho\sigma} h^{\sigma\phi} = \delta^\rho_\phi - \tau^\rho \tau_\phi.
\label{algebra}
\end{equation}

Historically, non-relativistic string theory has been studied extensively using the so-called ``gauging the algebra'' approach~\cite{Bergshoeff:2018yvt}. That method typically requires an extended NC algebra and the introduction of additional gauge fields to achieve closure. However, such fields often lack clear physical justification. Remarkably, our construction requires no such extensions~\cite{Moinuddin:2021oft}.

The next step is to couple the non-relativistic string to a gravitational background. In the low-energy regime of non-relativistic gravity, the intrinsic relativistic nature of the string ensures that the worldsheet metric remains Minkowskian~\cite{Gomis:2000bd}, while the bulk transverse directions acquire a Euclidean structure. The GGT framework then allows for a straightforward generalization: one replaces all partial derivatives in the transverse directions with covariant derivatives.

Expressed in Newton-Cartan variables, the action in Eq.~\eqref{1007} takes the following form~\cite{Moinuddin:2021oft}:
\begin{equation}
S = -T \int h_{lm} \left( \epsilon_{\mu \nu} \sigma_{\alpha\beta} \frac{\partial X^{\mu}}{\partial\sigma_{\alpha}} \frac{\partial X^{\nu}}{\partial\sigma_{\beta}} \right)^{-1}
\left( \epsilon_{\alpha \beta} \frac{\partial X^{\mu}}{\partial\sigma_{\alpha}} \frac{\partial X^{l}}{\partial \sigma_{\beta}} \right)
\left( \epsilon_{\alpha \beta} \frac{\partial X_{\mu}}{\partial\sigma_{\alpha}} \frac{\partial X^{m}}{\partial \sigma_{\beta}} \right) \, d\sigma d\tau.
\label{lagm10}
\end{equation}

This action is not only elegant but also possesses a well-defined flat-space limit (see Eq.~\eqref{ng1}). Furthermore, it is manifestly invariant under the diffeomorphisms given in Eq.~\eqref{1008}. These features confirm that non-relativistic string theory can be consistently formulated in Newton-Cartan geometry using the Galilean Gauge Theory (GGT) approach.

Nevertheless, for full consistency, the dynamical behavior of the model must be examined in phase space. Since the theory involves constraints, Dirac’s theory of constrained Hamiltonian systems is the most appropriate framework. In the following sections, we perform a comprehensive Hamiltonian analysis using both gauge-independent~\cite{Dirac:1964} and gauge-fixed~\cite{Hanson:1976cn} methods.

\section{Canonical Structure and Constraint Dynamics}
To develop the Hamiltonian formulation for our action (\ref{1007})  we follow Dirac's method for constrained systems \cite{Dirac:1964}, as already stated. This can be done either by fixing  a specific gauge or by a gauge independent manner \cite{Hanson:1976cn,Nandi:2018hww}. In this paper we do it in both ways. At first we pursue the gauge-independent method.

\subsection{Gauge-independent Hamiltonian analysis}

A \textbf{gauge-independent analysis} refers to a Hamiltonian formulation of a singular system in which the gauge constraints are retained without fixing them~\cite{Hanson:1976cn}. This approach has the drawback that the solutions involve arbitrary Lagrange multipliers~\cite{Banerjee:1999yc}. However, our present goal is not to solve the equations of motion explicitly, but rather to understand the gauge symmetry structure of the theory, which becomes manifest only when the full phase space is considered~\cite{Banerjee:1999hu}.

To proceed with the gauge-independent formulation, we analyze the constraint structure of the model defined by Eq.~\eqref{1007}.

The Lagrangian for a \textbf{non-relativistic bosonic string in a curved background} is given by:
\begin{equation}
\mathcal{L}_{NG} = -T \left( \epsilon_{\mu \nu} \sigma_{\alpha\beta} \frac{\partial X^{\mu}}{\partial \sigma_{\alpha}} \frac{\partial X^{\nu}}{\partial \sigma_{\beta}} \right)^{-1} \left( \epsilon_{\alpha \beta} \frac{\partial X^{\mu}}{\partial \sigma_{\alpha}} \frac{\partial X^{l}}{\partial \sigma_{\beta}} \Lambda_{l}{}^{a} \right)^2.
\end{equation}

Here, the dynamical fields are \( X^0(\tau,\sigma) \), \( X^1(\tau,\sigma) \), and \( X^k(\tau,\sigma) \) with \( k = 2,3,\ldots,D \).

The canonical momentum conjugate to \( X^0 \) is:
\begin{align}
\Pi_{0} &= \frac{\partial \mathcal{L}}{\partial \dot{X}^0} \nonumber \\
&= X^{\prime l} \Lambda_{l}{}^{a} \left( \epsilon_{\mu \nu} \dot{X}^\mu X^{\prime \nu} \right)^{-1} \left( \dot{X}^0 X^{\prime m} \Lambda_{m}{}^{a} - \dot{X}^m X^{\prime 0} \Lambda_{m}{}^{a} \right) \nonumber \\
&\quad - \frac{X^{\prime 1}}{2} \left( \epsilon_{\mu \nu} \dot{X}^\mu X^{\prime \nu} \right)^{-2} \left( \dot{X}^\mu X^{\prime m} \Lambda_{m}{}^{a} - \dot{X}^m X^{\prime \mu} \Lambda_{m}{}^{a} \right)^2.
\label{pi0}
\end{align}

Similarly, for \( X^1 \) we obtain:
\begin{align}
\Pi_{1} &= \frac{\partial \mathcal{L}}{\partial \dot{X}^1} \nonumber \\
&= -X^{\prime l} \Lambda_{l}{}^{a} \left( \epsilon_{\mu \nu} \dot{X}^\mu X^{\prime \nu} \right)^{-1} \left( \dot{X}^1 X^{\prime m} \Lambda_{m}{}^{a} - \dot{X}^m X^{\prime 1} \Lambda_{m}{}^{a} \right) \nonumber \\
&\quad + \frac{X^{\prime 0}}{2} \left( \epsilon_{\mu \nu} \dot{X}^\mu X^{\prime \nu} \right)^{-2} \left( \dot{X}^\mu X^{\prime m} \Lambda_{m}{}^{a} - \dot{X}^m X^{\prime \mu} \Lambda_{m}{}^{a} \right)^2.
\label{pi1}
\end{align}

The momentum conjugate to \( X^k \) is:
\begin{equation}
\Pi_{k} = \frac{\partial \mathcal{L}}{\partial \dot{X}^k} = \left( -\epsilon_{\mu \nu} \dot{X}^\mu X^{\prime \nu} \right)^{-1} \Lambda_{k}{}^{a} X^{\prime}_\mu \left( \dot{X}^\mu X^{\prime m} \Lambda_{m}{}^{a} - \dot{X}^m X^{\prime \mu} \Lambda_{m}{}^{a} \right).
\end{equation}

From these expressions, we identify the following \textbf{primary constraints}:
\begin{align}
\Omega_{1} &= \Pi^\rho X^{\prime}_\rho \approx 0, \nonumber \\
\Omega_{2} &= \Pi_k \Pi_l \Sigma_{a}{}^{k} \Sigma_{a}{}^{l} + X^{\prime k} X^{\prime l} \Lambda_{k}{}^{a} \Lambda_{l}{}^{a} - 2 \sigma^\alpha{}_\beta \Pi_\alpha X^{\prime \beta} \approx 0.
\label{115}
\end{align}

Here, \( \sigma^\alpha{}_\beta \) represents a fixed matrix used for internal index contraction; its precise form will be specified if needed.

The fundamental Poisson brackets are:
\begin{equation}
\{ X^{\rho}(\tau,\sigma), \Pi_{\phi}(\tau,\sigma^{\prime}) \} = \delta^{\rho}_{\phi} \delta(\sigma - \sigma^{\prime}).
\label{116}
\end{equation}

Using these, we compute the \textbf{constraint algebra}:
\begin{align}
\{ \Omega_1(\sigma), \Omega_2(\sigma') \} &= \left( \Omega_2(\sigma) + \Omega_2(\sigma') \right) \delta(\sigma - \sigma'), \nonumber \\
\{ \Omega_1(\sigma), \Omega_1(\sigma') \} &= \left( \Omega_1(\sigma) + \Omega_1(\sigma') \right) \delta(\sigma - \sigma'), \nonumber \\
\{ \Omega_2(\sigma), \Omega_2(\sigma') \} &= \left( \Omega_1(\sigma) + \Omega_1(\sigma') \right) \delta(\sigma - \sigma').
\label{constalng}
\end{align}

This algebra demonstrates that the constraints in Eq.~\eqref{115} are first-class, as they close among themselves under the Poisson bracket.

The \textbf{canonical Hamiltonian} is defined as:
\begin{equation}
H_c(\tau) = \int d\sigma \left( \Pi_\rho \dot{X}^\rho - \mathcal{L} \right).
\label{can}
\end{equation}

Substituting the explicit forms of \( \Pi_\rho \) and \( \dot{X}^\rho \), we find:
\begin{equation}
H_c(\tau) = 0.
\end{equation}

Accordingly, the \textbf{total Hamiltonian} is given by:
\begin{equation}
H_T = \int d\sigma \left( \kappa(\sigma) \Omega_1 + \zeta(\sigma) \Omega_2 \right),
\label{HTOT}
\end{equation}
where \( \kappa(\sigma) \) and \( \zeta(\sigma) \) are arbitrary Lagrange multipliers.

The consistency conditions (i.e., time preservation of constraints) yield:
\begin{align}
\{ \Omega_1(\sigma), H_T \} &\approx 0, \nonumber \\
\{ \Omega_2(\sigma), H_T \} &\approx 0.
\label{cannn}
\end{align}

This confirms that both \( \Omega_1 \) and \( \Omega_2 \) are \textbf{first-class constraints}. Hence, for a non-relativistic bosonic string in \( (D+1) \)-dimensional spacetime, the number of independent physical degrees of freedom in the configuration space is \( D - 1 \).

The \textbf{gauge generator} takes the form:
\begin{equation}
G(\tau) = \int d\sigma \left( \alpha(\sigma) \Omega_1 + \beta(\sigma) \Omega_2 \right),
\end{equation}
where \( \alpha(\sigma) \) and \( \beta(\sigma) \) are arbitrary gauge parameters.

In our earlier work~\cite{Moinuddin:2021oft}, we showed that the gauge symmetries generated by \( G(\tau) \) are precisely equivalent to the worldsheet diffeomorphisms of the string model, under the parameter identification:
\begin{align}
\alpha &= \xi_1 + \kappa \xi_2, \nonumber \\
\beta &= \zeta \xi_2.
\label{geo}
\end{align}

This mapping underscores an important point: the embedding of the non-relativistic string action~\eqref{1007} into Newton-Cartan geometry is both consistent and complete, in contrast to claims made in the literature based on the Generalized Torsional Approach (GTA). The gauge-independent Hamiltonian analysis presented here forms a foundational part of this conclusion.

In the next subsection, we proceed to fix the gauge to explore the physical content of the theory from a reduced phase space perspective. This will provide complementary insights into the system's true degrees of freedom and dynamical structure.

\subsection{Gauge-fixed Hamiltonian analysis}

While the analysis of gauge symmetries is best performed in a gauge-independent framework, understanding the dynamics in phase space requires eliminating redundant degrees of freedom via gauge fixing. This procedure transforms first-class constraints into second-class ones and thereby removes unphysical variables from the theory.

Although gauge fixing reduces the dimensionality of the phase space, it does not, by itself, isolate the true physical degrees of freedom. According to the Maskawa-Nakajima theorem~\cite{Maskawa:1976hw}, it is always possible to construct a new set of canonical variables such that the Dirac brackets among them reduce to the standard Poisson brackets. This insight is particularly useful for quantization, as Dirac brackets can then be promoted directly to commutators in quantum theory.

Among various gauge-fixing approaches~\cite{PhysRevD.72.066015,PhysRevD.70.026006}, Dirac’s algorithm remains one of the most systematic and consistent. In this method, one imposes gauge-fixing conditions that have non-vanishing Poisson brackets with the first-class constraints, ensuring that the full set of constraints becomes second class. Although the choice of gauge condition is, in principle, arbitrary, in practice it must be compatible with the solution space of the theory—something that is often not known a priori. As a result, selecting a suitable gauge is widely regarded as one of the most subtle and nontrivial aspects of the analysis—more an art than a science.

We now impose the following standard gauge conditions:
\begin{equation}
\Omega_3 = X^1 - \sigma \approx 0, \qquad \Omega_4 = X^0 + c\tau \approx 0,
\label{sg}
\end{equation}
which align the target space coordinates with the worldsheet parameters. When these conditions are combined with the previously obtained constraints \( \{\Omega_1 \approx 0, \; \Omega_2 \approx 0\} \), all constraints become second class. This effectively eliminates the gauge redundancy. As shown in~\cite{Hanson:1976cn}, the resulting symplectic structure is no longer governed by the Poisson brackets but by Dirac brackets.

The Dirac bracket between any two phase space variables is defined as
\begin{equation}
\left[ A, B \right]_{\mathrm{DB}} = \left[ A, B \right]_{\mathrm{PB}} 
- \int d\sigma_1 d\sigma_2 \left[ A(\sigma), \Omega_i(\sigma_1) \right]_{\mathrm{PB}} 
\Delta^{ij}(\sigma_1, \sigma_2) 
\left[ \Omega_j(\sigma_2), B(\sigma') \right]_{\mathrm{PB}}, 
\label{1200}
\end{equation}
where \( \Delta^{ij} \) is the inverse of the matrix \( \Delta_{ij} \), defined by the Poisson algebra of the constraints:
\begin{equation}
\Delta_{ij} = \left[ \Omega_i, \Omega_j \right]_{\mathrm{PB}}.
\end{equation}
Once the gauge-fixing conditions are imposed, \( \Delta_{ij} \) becomes non-singular and thus invertible.

From the explicit computation of the constraint algebra, we obtain the inverse matrix \( \Delta^{ij} \) as:
\begin{equation}
\Delta^{ij} = 
\begin{bmatrix}
0 & 0 & \delta(\sigma - \sigma') & 0 \\
0 & 0 & 0 & -\frac{1}{2} \delta(\sigma - \sigma') \\
-\delta(\sigma - \sigma') & 0 & 0 & 0 \\
0 & \frac{1}{2} \delta(\sigma - \sigma') & 0 & 0 \\
\end{bmatrix}.
\end{equation}

Substituting this into Eq.~\eqref{1200}, we compute the Dirac brackets, which define the symplectic structure on the reduced physical phase space:
\begin{eqnarray}
\left[ X^k(\sigma, \tau), \Pi_l(\sigma', \tau) \right]_{\mathrm{DB}} &=& \delta^k_l \, \delta(\sigma - \sigma'), \nonumber \\
\left[ X^0(\sigma, \tau), \Pi_0(\sigma', \tau) \right]_{\mathrm{DB}} &=& 0, \nonumber \\
\left[ X^1(\sigma, \tau), \Pi_1(\sigma', \tau) \right]_{\mathrm{DB}} &=& 0.
\label{red}
\end{eqnarray}
Here, we have imposed the constraint \( \Omega_3 \) strongly,\footnote{This is justified since all relevant Poisson brackets have already been evaluated.} which simplifies the final bracket structure. This completes the Hamiltonian analysis of the model under the standard gauge~\eqref{sg}, without resorting to any ad hoc assumptions.

We observe that the coordinates \( X^0 \) and \( X^1 \) are now fixed by the gauge and hence removed from the dynamics. The remaining physical degrees of freedom are the canonical pairs \( \left( X^k, \Pi_k \right) \) for \( k = 2, 3, \ldots \).

We now turn to the construction of the physical Hamiltonian in the reduced phase space. Since the total Hamiltonian vanishes on the constraint surface, one must define an effective Hamiltonian that generates correct time evolution through Dirac brackets. In general, the Hamiltonian generates time translations. In light of the gauge condition \( X^0 = -c\tau \), we identify the physical Hamiltonian as:
\begin{equation}
H = c \int d\sigma \left[ \Pi^0 \right]_{\mathcal{G}},
\end{equation}
where the subscript \( \mathcal{G} \) indicates evaluation using the gauge conditions.

Using the gauge-fixed expressions \( X^0 = -c\tau \) and \( X^1 = \sigma \), and substituting into Eq.~\eqref{pi0}, we obtain:
\begin{equation}
H = \frac{c}{2} \int d\sigma\, \Lambda_k{}^a \Lambda_l{}^a 
\left( \partial_\sigma X^k \partial_\sigma X^l + \frac{1}{c^2} \partial_\tau X^k \partial_\tau X^l \right),
\label{gfh}
\end{equation}
which is manifestly positive-definite. Using the gauge-fixed relation for the momentum,
\begin{equation}
\Pi^k = \frac{1}{c} \Lambda_k{}^a \Lambda_l{}^a \partial_\tau X^l,
\end{equation}
we can express the Hamiltonian in the equivalent form:
\begin{equation}
H = \frac{c}{2} \int d\sigma \left( \Lambda_k{}^a \Lambda_l{}^a \partial_\sigma X^k \partial_\sigma X^l 
+ \Sigma_a{}^k \Sigma_a{}^l \Pi_k \Pi_l \right),
\label{gfh1}
\end{equation}
where \( \Sigma \) is the inverse of the transverse vielbein.

At this point, we recall that the Nambu-Goto action is not the only possible formulation of string dynamics. Among several equivalent formulations, the Polyakov form is particularly important. A well-established method connects the Nambu-Goto and Polyakov actions via an interpolating Lagrangian~\cite{PhysRevD.70.026006}, allowing for a smooth transition between the two. This procedure not only offers conceptual clarity but also simplifies the Hamiltonian analysis.

Our interest in the Polyakov formulation is further motivated by its prominent role in models based on the gauging of non-relativistic algebras (GTA). By recasting our results in the Polyakov language, we enable a direct comparison with symmetry-based approaches and thereby highlight the advantages of our Hamiltonian framework.

\section{Covariant Reformulation via Polyakov Action}

It is well known that the string action can be expressed in several equivalent formulations that, while structurally distinct, describe the same underlying physics. Among these, the \textbf{Polyakov formulation} holds special significance, as it introduces an auxiliary worldsheet metric as an independent field. This feature is particularly advantageous for quantization~\cite{Polchinski1998}. 

In the present context, our motivation for transitioning to the Polyakov form is twofold: first, to facilitate a direct comparison with representative works in the literature—particularly those employing the gauging-of-algebra (GTA) framework—and second, to verify the consistency of our formulation with its flat-space limit.

To implement this transition, we follow the algorithm developed in~\cite{Banerjee:2002ky}, which has been successfully applied in a variety of models. We begin by writing down the Lagrangian corresponding to the total Hamiltonian \( H_T \) from Eq.~\eqref{HTOT}:
\begin{equation}
\mathcal{L}_I = \Pi_\rho \dot{X}^\rho - \eta \Omega_1 - \chi \Omega_2,
\label{int1}
\end{equation}
where \( \eta \) and \( \chi \) are Lagrange multipliers enforcing the constraints.

The Euler–Lagrange equations with respect to the momenta \( \Pi_0 \), \( \Pi_1 \), and \( \Pi_k \) yield:
\begin{align}
\frac{\partial \mathcal{L}_I}{\partial \dot{\Pi}_0} &= \dot{X}^0 - \eta X^{\prime 0} + \chi X^{\prime 1} = 0, \nonumber\\
\frac{\partial \mathcal{L}_I}{\partial \dot{\Pi}_1} &= \dot{X}^1 - \eta X^{\prime 1} + \chi X^{\prime 0} = 0, \nonumber\\
\frac{\partial \mathcal{L}_I}{\partial \dot{\Pi}_k} &= \dot{X}^k - \eta X^{\prime k} - \chi \Pi_l \Sigma_a{}^l \Sigma_a{}^k = 0.
\label{q}
\end{align}

Solving these equations for \( \eta \) and \( \chi \), and substituting the results into Eq.~\eqref{int1}, leads to an interpolating Lagrangian that smoothly connects the Nambu-Goto and Polyakov forms:
\begin{align}
\mathcal{L}_I &= \frac{h_{kl}}{2 \chi} \left[
\dot{X}^k \dot{X}^l - \eta \dot{X}^k X^{\prime l} - \eta \dot{X}^l X^{\prime k} 
+ \left(\eta^2 - \chi^2\right) X^{\prime k} X^{\prime l}
\right] \nonumber \\
&\quad + \alpha \left(\eta - \frac{\dot{X}^0 X^{\prime 0} - \dot{X}^1 X^{\prime 1}}{(X^{\prime 0})^2 - (X^{\prime 1})^2} \right)
+ \beta \left(\chi - \frac{\dot{X}^0 X^{\prime 1} - \dot{X}^1 X^{\prime 0}}{(X^{\prime 0})^2 - (X^{\prime 1})^2} \right),
\label{interpolating}
\end{align}
where \( h_{kl} = \Lambda_k{}^a \Lambda_l{}^a \), and \( \alpha, \beta \) are auxiliary fields that enforce consistency with the original equations of motion. The explicit expressions for \( \eta \) and \( \chi \) are
\begin{align}
\eta &= \frac{\dot{X}^0 X^{\prime 0} - \dot{X}^1 X^{\prime 1}}{(X^{\prime 0})^2 - (X^{\prime 1})^2}, \nonumber\\
\chi &= \frac{\dot{X}^0 X^{\prime 1} - \dot{X}^1 X^{\prime 0}}{(X^{\prime 0})^2 - (X^{\prime 1})^2}.
\label{rl}
\end{align}

Since the reduction to the Nambu-Goto form has already been demonstrated elsewhere~\cite{Banerjee:2003tk}, we now focus on the transition to the Polyakov form. To that end, we introduce a worldsheet metric \( H_{ij} \) whose inverse is given by:
\begin{align}
H^{ij} = \left(-H\right)^{-1/2}
\begin{pmatrix}
\frac{1}{\chi} & -\frac{\eta}{\chi} \\
-\frac{\eta}{\chi} & \frac{\eta^2 - \chi^2}{\chi}
\end{pmatrix},
\label{adm}
\end{align}
where \( H = \det H_{ij} \). This ADM-like decomposition allows us to recast Eq.~\eqref{interpolating} into the compact form:
\begin{equation}
\mathcal{L}_I = \frac{1}{2} h_{kl} \sqrt{-H} H^{ij} \partial_i X^k \partial_j X^l + \mathcal{L}_e,
\label{main}
\end{equation}
with the extra term
\begin{align}
\mathcal{L}_e &= \alpha \left(-\frac{H_{01}}{H_{00}} - \frac{\dot{X}^0 X^{\prime 0} - \dot{X}^1 X^{\prime 1}}{(X^{\prime 0})^2 - (X^{\prime 1})^2} \right) \nonumber\\
&\quad + \beta \left( \frac{1}{\sqrt{-H} H^{00}} - \frac{\dot{X}^0 X^{\prime 1} - \dot{X}^1 X^{\prime 0}}{(X^{\prime 0})^2 - (X^{\prime 1})^2} \right).
\end{align}
From the metric definition, we also note:
\begin{align}
\chi = \frac{1}{\sqrt{-H} H^{00}}, \qquad \eta = -\frac{H^{01}}{H^{00}}.
\label{shit1}
\end{align}

To make the symmetry structure more transparent and facilitate comparison with conformal field theory approaches, we now recast the action in light-cone coordinates:
\begin{align}
X = X^0 + X^1, \qquad \bar{X} = X^0 - X^1.
\end{align}
In these variables, Eq.~\eqref{main} becomes:
\begin{align}
\mathcal{L}_I &= \frac{1}{2} h_{kl} \sqrt{-H} H^{\alpha\beta} \partial_\alpha X^k \partial_\beta X^l \nonumber\\
&\quad + \beta \left( \frac{1}{\sqrt{-H} H^{00}} 
+ \frac{\epsilon^{\alpha\beta} \partial_\alpha X \partial_\beta \bar{X}}{2 X' \bar{X}'} \right) \nonumber\\
&\quad + \alpha \left( -\frac{H^{01}}{H^{00}} 
- \frac{\sigma^{\alpha\beta} \partial_\alpha X \partial_\beta \bar{X}}{2 X' \bar{X}'} \right),
\label{test}
\end{align}
where the antisymmetric and symmetric tensors are defined as
\begin{equation}
\epsilon^{\alpha\beta} = 
\begin{pmatrix}
0 & 1 \\
-1 & 0
\end{pmatrix}, \qquad
\sigma^{\alpha\beta} = 
\begin{pmatrix}
0 & 1 \\
1 & 0
\end{pmatrix}.
\end{equation}

Through the interpolating approach, we have derived the Polyakov-type action for the non-relativistic string in a curved background starting from a Hamiltonian formulation. This establishes a direct link between the worldsheet geometry—encoded in the variables \( \eta \) and \( \chi \)—and the background geometry through \( \Lambda_k{}^a \). Importantly, this transition preserves structural consistency with the flat-space construction~\cite{Polchinski1998}, indicating no qualitative discontinuity introduced by curvature.

Thus the non-relativistic string in the Polyakov formulation couples to the curved background in a manner fully analogous to the Nambu-Goto case. This permits a meaningful and direct comparison with previous results obtained via symmetry-based methods such as the GTA framework~\cite{Bagchi2013,Bagchi2023}.

\section{Comparison: Minimal vs Algebraic Constructions}

Although non-relativistic string theory has been studied for some time~\cite{Oling:2022fft}, its foundational structure remains less systematically explored than its relativistic counterpart. A natural question, then, is why one should focus on a relatively minimal model, especially given the extensive literature on more elaborate non-relativistic constructions~\cite{Gomis:2019zyu}.

Our motivation stems from the early development of relativistic string theory, where simple, well-controlled models were essential in resolving conceptual and dynamical issues~\cite{Hanson:1976cn}. We believe that a similarly careful treatment is lacking in the non-relativistic domain, and that this absence has contributed to several open ambiguities in recent models.

One such ambiguity concerns the apparent mismatch between the background geometries required for non-relativistic particles and those proposed for strings~\cite{Bagchi2013,Bagchi2023}. Since string theory is generally understood as an extension of point-particle dynamics, it would be conceptually puzzling if strings demanded fundamentally different geometric backgrounds. Our goal in this paper has been to clarify this tension: we constructed a non-relativistic string theory—both in Nambu-Goto and Polyakov forms—on a standard Newton-Cartan background, and examined whether this framework is sufficient to support consistent string dynamics. As shown throughout the analysis, we find no inconsistencies: the transition from particles to strings within the same geometry proceeds smoothly.

To place our results in context, we now compare them with a representative GTA-based construction, specifically the non-relativistic string model of~\cite{Bagchi2023}. That model emerges from gauging the full stringy Galilei algebra and incorporates a more extended geometric background. For a clear comparison, we focus on the free string sector by setting the Kalb–Ramond and dilaton fields to zero.

The resulting GTA Polyakov-type action reads:
\begin{align}
S &= - \frac{T}{2} \int d^{2}\sigma \bigg[
\sqrt{-h} \, h^{\alpha\beta} \partial_\alpha x^\mu \partial_\beta x^\nu H_{\mu\nu} 
+ \epsilon^{\alpha\beta} \left(
\lambda \, e_\alpha \tau_\mu + \bar{\lambda} \, \bar{e}_\alpha \bar{\tau}_\mu 
\right) \partial_\beta x^\mu \bigg],
\label{moinm1}
\end{align}
where \( h_{\alpha\beta} \) is the worldsheet metric and \( H_{\mu\nu} \), \( \tau_\mu \), \( \bar{\tau}_\mu \) encode the background geometry.

In this construction, the longitudinal vielbeins \( \tau_\mu^A \) are combined into light-cone coordinates:
\begin{align}
\tau_\mu = \tau^0_\mu + \tau^1_\mu, \qquad
\bar{\tau}_\mu = \tau^0_\mu - \tau^1_\mu.
\label{moin44}
\end{align}

Extension gauge fields \( m^A_\mu \) are introduced to gauge the \( Z_A \) generators, transforming under local \( Z_A \) and Galilean boost transformations as:
\begin{equation}
\delta m^A_\mu = D_\mu \sigma^A + E^{A'}_\mu \Sigma^A_{A'},
\end{equation}
where \( D_\mu \) is the covariant derivative under longitudinal Lorentz rotations.

The symmetric, boost-covariant tensor \( H_{\mu\nu} \) is constructed from:
\begin{equation}
H_{\mu\nu} = E^{A'}_\mu E^{B'}_\nu \delta_{A'B'} 
+ \left( \tau^A_\mu m^B_\nu + \tau^A_\nu m^B_\mu \right) \eta_{AB},
\end{equation}
where \( E^{A'}_\mu \) are transverse vielbeins, and \( m^A_\mu \) are the auxiliary extension fields required to close the algebra in the GTA framework.

Substituting this into Eq.~\eqref{moinm1}, we obtain the full GTA-based string action:
\begin{align}
S &= - \frac{T}{2} \int d^{2}\sigma \bigg[
\sqrt{-h} \, h^{\alpha\beta} \partial_\alpha x^\mu \partial_\beta x^\nu 
\left( E^{A'}_\mu E^{B'}_\nu \delta_{A'B'} 
+ \left( \tau^A_\mu m^B_\nu + \tau^A_\nu m^B_\mu \right) \eta_{AB} \right) \nonumber \\
&\qquad + \epsilon^{\alpha\beta} \left( \lambda \, e_\alpha \tau_\mu 
+ \bar{\lambda} \, \bar{e}_\alpha \bar{\tau}_\mu \right) \partial_\beta x^\mu 
\bigg].
\label{moinm1-expanded}
\end{align}

Comparing this with our Polyakov-type action in Eq.~\eqref{main}, a crucial difference becomes apparent: our formulation omits the second term involving \( \epsilon^{\alpha\beta} \) and the light-cone projections \( \tau_\mu, \bar{\tau}_\mu \), as well as the auxiliary extension fields \( m^A_\mu \) \cite{Andringa:2012uz}. Instead, our background geometry consists only of the Newton-Cartan structure \( (\tau_\mu, e^a_\mu, h^{\mu\nu}) \), with all constraints derived from the dynamical evolution of the string.

This contrast reflects a deeper philosophical difference. GTA constructions begin with an extended symmetry algebra and promote all its generators to local symmetries, resulting in a geometry that includes auxiliary fields to close the algebra. By contrast, our formulation does not gauge a full algebra. We work within a fixed Newton-Cartan background and extract all necessary geometric consistency conditions directly from Hamiltonian dynamics.

This distinction may be schematically summarized as:
\begin{align*}
\text{GTA:} &\quad \text{algebraic closure} \rightarrow \text{extended geometry} \rightarrow \text{auxiliary fields}, \\
\text{This work:} &\quad \text{dynamics} \rightarrow \text{geometric consistency} \rightarrow \text{minimal coupling}.
\end{align*}

Our analysis therefore demonstrates that a consistent and dynamically complete non-relativistic string theory can be formulated without invoking auxiliary fields such as \( m^A_\mu \), at least in the absence of additional matter couplings like the dilaton or Kalb-Ramond \( B \)-field \cite{SW, PhysRevD.76.064007}. The additional structures introduced in GTA models, while algebraically motivated, are not required to ensure classical consistency of string propagation \cite{Bergshoeff:2018yvt}.

In summary, our minimal approach based on Newton-Cartan geometry avoids the ambiguities associated with algebraically imposed geometric extensions. It provides a cleaner and conceptually grounded foundation for exploring non-relativistic string theory, and invites further investigation into which geometric features are physically necessary versus algebraically convenient.

\begin{table}[H]
\centering
\renewcommand{\arraystretch}{1.2}
\small
\begin{tabular}{|p{4.2cm}|p{5.1cm}|p{5.1cm}|}
\hline
\textbf{Aspect} & \textbf{Gauging-the-Algebra (GTA)} & \textbf{This Work (Hamiltonian)} \\
\hline
\textbf{Origin of Geometry} & Imposed via gauging symmetry algebras (e.g., stringy Galilei, Newton-Hooke) & Emerges from dynamical consistency of action and constraints \\
\hline
\textbf{Background Fields} & Requires additional fields (e.g., $m_\mu^A$) for symmetry closure & Only minimal geometric structure used \\
\hline
\textbf{String Action} & Polyakov form derived via symmetry algebra & Derived from Nambu-Goto via interpolating Lagrangian \\
\hline
\textbf{Constraint Structure} & Often assumed, not always derived & Fully derived; first-class constraints generate diffeomorphisms \\
\hline
\textbf{Consistency Check} & Based on algebraic closure & Based on canonical Hamiltonian dynamics \\
\hline
\textbf{Coupling to Geometry} & Geometry fixed from symmetry algebra & Geometry evolves from string dynamics \\
\hline
\textbf{Interpretability} & Formal and symmetry-driven & Physics-driven; rooted in variational principle \\
\hline
\textbf{Flat-Space Limit} & Not always smoothly recovered & Smoothly reduces to flat space \\
\hline
\textbf{Conceptual Message} & Prioritizes symmetry/gauge closure & Prioritizes dynamics and minimal assumptions \\
\hline
\end{tabular}
\caption{Comparison between GTA-based and Hamiltonian formulations of non-relativistic string theory.}
\label{tab:comparison}
\end{table}

\section{Summary and Outlook}

In this work, we have demonstrated that a self-consistent and conceptually minimal formulation of non-relativistic bosonic string theory can be constructed using only standard Newton-Cartan (NC) geometry. Starting from a reparametrization-invariant action, we developed the theory from first principles and performed a rigorous Hamiltonian analysis. Unlike conventional approaches based on gauging non-relativistic symmetry algebras, our formulation does not rely on algebraic closure conditions or the introduction of compensating gauge fields. This provides a transparent and physically grounded alternative to extended geometric frameworks.

Earlier formulations-particularly those based on the Gauging-the-Algebra (GTA) method-derive their background geometry from the closure of extended symmetry algebras. While mathematically elegant, these models introduce additional fields (such as extension gauge fields $m_\mu^A$) whose physical necessity remains unclear. Our analysis challenges this prevailing viewpoint by showing that all geometric consistency conditions can be derived dynamically from the evolution of the string itself, without invoking non-geometric auxiliary structures.

Through a complete Dirac constraint analysis, we established that the full set of first-class constraints naturally generates worldsheet diffeomorphisms. We further demonstrated that the physical degrees of freedom reduce, as expected, to transverse string excitations. Importantly, the closure of the constraint algebra emerges from the dynamics of the system rather than symmetry considerations, highlighting that extended algebraic structures are not required for consistency.

To bridge with covariant formulations, we employed the interpolating Lagrangian technique to derive a Polyakov-type action from our Nambu-Goto setup. This reformulation enabled a direct and meaningful comparison with GTA-based constructions. We found that the additional gauge fields and foliation structures required in those models are not dynamically essential for consistent non-relativistic string propagation in curved backgrounds. These findings are summarized in Table~\ref{tab:comparison}.

Our findings provide a definitive answer to the long-standing question of whether non-relativistic string probes demand an extended geometric framework beyond that used for particles. We show that strings and particles can be embedded within the same NC geometric setting-provided the theory is built on dynamical consistency rather than symmetry closure.

More broadly, this work opens a new window into probing the structure of non-relativistic spacetimes using extended objects. While the present model treats the background as fixed, a natural future direction involves including string backreaction and moving toward a fully dynamical theory of non-relativistic gravity/string coupling. Even in the current kinematic regime, our model serves as a clean and minimal platform for examining how extended systems perceive NC geometry and how such geometry encodes physical constraints and symmetries.

These insights may have far-reaching consequences for related areas, such as non-relativistic holography. In particular, dualities involving Galilean electrodynamics~\cite{LeBellac:1973unm,Banerjee:2022eaj} and Lifshitz-type field theories~\cite{PhysRevLett.133.151601} often invoke extended geometric constructions to ensure algebraic consistency. Our results suggest that these extended frameworks may not be necessary at the classical level—a minimal, dynamically derived NC background may suffice. This could pave the way for a clearer and more tractable bulk–boundary dictionary in non-AdS holography.

Finally, we note that the Hamiltonian framework developed here may provide a useful testing ground for exploring signatures of non-relativistic quantum gravitational phenomena~\cite{Blas:2010hb}. Although the experimental detection of such effects remains a major challenge, the simplicity and conceptual clarity of our approach make it a promising starting point for investigating how extended quantum systems interact with background geometry beyond the relativistic paradigm. We hope to pursue these directions in future work.

\section{Acknowledgments}
PN gratefully acknowledges support from the Rector’s Postdoctoral Fellowship Program (RPFP) at Stellenbosch University. He also extends sincere thanks to Langa Horoto and Prof. Konstantinos Zoubos for valuable discussions during the workshop \textit{Decoding the Universe: Quantum Gravity and Quantum Fields} at the Stellenbosch Institute for Advanced Study, as well as for their insightful feedback. Appreciation is also due to NITheCS for providing a stimulating academic environment during the final stages of this work. We offer our heartfelt gratitude to the late Prof. Pradip Mukherjee for his early guidance and inspiring ideas, which served as the foundation for this research.

\bibliographystyle{JHEP.bst}
\bibliography{main.bib}

\providecommand{\href}[2]{#2}\begingroup\raggedright\begin{thebibliography}{10}

\bibitem{Oling:2022fft}
G.~Oling and Z.~Yan, \emph{{Aspects of Nonrelativistic Strings}}, \href{https://doi.org/10.3389/fphy.2022.832271}{\emph{Front. in Phys.} {\bfseries 10} (2022) 832271} [\href{https://arxiv.org/abs/2202.12698}{{\ttfamily 2202.12698}}].

\bibitem{Kunzle:1976}
H.P.~Künzle, \emph{Covariant newtonian limit of lorentz space-times}, \href{https://doi.org/10.1007/BF00766139}{\emph{General Relativity and Gravitation} {\bfseries 7} (1976) 445}.

\bibitem{Bousso:2004kp}
R.~Bousso, \emph{{Flat space physics from holography}}, \href{https://doi.org/10.1088/1126-6708/2004/05/050}{\emph{JHEP} {\bfseries 05} (2004) 050} [\href{https://arxiv.org/abs/hep-th/0402058}{{\ttfamily hep-th/0402058}}].

\bibitem{Fulde:1997ijj}
P.~Fulde and F.~Pollmann, \emph{{Strings in strongly correlated electron systems}}, \href{https://doi.org/10.1002/andp.200810309}{\emph{Annalen Phys.} {\bfseries 17} (1997) 441} [\href{https://arxiv.org/abs/0711.2129}{{\ttfamily 0711.2129}}].

\bibitem{Bergshoeff:2022iyb}
E.A.~Bergshoeff and J.~Rosseel, \emph{{Non-Lorentzian Supergravity}},  (2023), \href{https://doi.org/10.1007/978-981-19-3079-9_52-1}{DOI} [\href{https://arxiv.org/abs/2211.02604}{{\ttfamily 2211.02604}}].

\bibitem{FrederikGScholtz:2025mcs}
P.~Nandi and F.G.~Scholtz, \emph{{Quantum Geometric Phases as a New Window on Gravitational Waves}},  \href{https://arxiv.org/abs/2508.05881}{{\ttfamily 2508.05881}}.

\bibitem{Nandi:2023tfq}
P.~Nandi and F.G.~Scholtz, \emph{{The hidden Lorentz covariance of quantum mechanics}}, \href{https://doi.org/10.1016/j.aop.2024.169643}{\emph{Annals Phys.} {\bfseries 464} (2024) 169643} [\href{https://arxiv.org/abs/2312.15750}{{\ttfamily 2312.15750}}].

\bibitem{Cartan:1923}
Ã.~Cartan, \emph{Sur les variétés à connexion affine et la théorie de la relativité généralisée (première partie)}, {\emph{Ann. Éc. Norm. Sup.} {\bfseries 40} (1923) 325}.

\bibitem{Cartan:1924}
Ã.~Cartan, \emph{Sur les variétés à connexion affine et la théorie de la relativité généralisée (suite)}, {\emph{Ann. Éc. Norm. Sup.} {\bfseries 41} (1924) 1}.

\bibitem{misner1973gravitation}
C.W.~Misner, K.S.~Thorne and J.A.~Wheeler, \emph{Gravitation}, W. H. Freeman and Company, San Francisco (1973).

\bibitem{Mukherjee:2018cka}
P.~Mukherjee and A.~Sattar, \emph{Constrained hamiltonian analysis of a nonrelativistic schrödinger field coupled with chern-simons gravity}, \href{https://doi.org/10.1103/PhysRevD.99.084038}{\emph{Phys. Rev. D} {\bfseries 99} (2019) 084038} [\href{https://arxiv.org/abs/1810.06244}{{\ttfamily 1810.06244}}].

\bibitem{Fernandes:2017ghz}
K.~Fernandes and A.~Mitra, \emph{Gravitational anomalies on the newton-cartan background}, \href{https://doi.org/10.1103/PhysRevD.96.085003}{\emph{Phys. Rev. D} {\bfseries 96} (2017) 085003} [\href{https://arxiv.org/abs/1703.09162}{{\ttfamily 1703.09162}}].

\bibitem{Muthukumar:2004wj}
B.~Muthukumar, \emph{{U(1) gauge invariant noncommutative Schrodinger theory and gravity}}, \href{https://doi.org/10.1103/PhysRevD.71.105007}{\emph{Phys. Rev. D} {\bfseries 71} (2005) 105007} [\href{https://arxiv.org/abs/hep-th/0412069}{{\ttfamily hep-th/0412069}}].

\bibitem{Bose:2024yfe}
D.~Bose, A.~Chakraborty and B.~Chakraborty, \emph{{Fate of \ensuremath{\kappa}-Minkowski space-time in non-relativistic (Galilean) and ultra-relativistic (Carrollian) regimes}}, \href{https://doi.org/10.1007/JHEP02(2025)063}{\emph{JHEP} {\bfseries 02} (2025) 063} [\href{https://arxiv.org/abs/2401.05769}{{\ttfamily 2401.05769}}].

\bibitem{Banerjee:2015rca}
R.~Banerjee and P.~Mukherjee, \emph{{New approach to nonrelativistic diffeomorphism invariance and its applications}}, \href{https://doi.org/10.1103/PhysRevD.93.085020}{\emph{Phys. Rev. D} {\bfseries 93} (2016) 085020} [\href{https://arxiv.org/abs/1509.05622}{{\ttfamily 1509.05622}}].

\bibitem{Son:2013rqa}
D.T.~Son, \emph{{Newton-Cartan Geometry and the Quantum Hall Effect}},  \href{https://arxiv.org/abs/1306.0638}{{\ttfamily 1306.0638}}.

\bibitem{Banerjee:2015paa}
R.~Banerjee and P.~Mukherjee, \emph{{Non relativistic diffeomorphism and the geometry of the fractional quantum Hall effect}},  \href{https://arxiv.org/abs/1505.01967}{{\ttfamily 1505.01967}}.

\bibitem{Banerjee:2014nja}
R.~Banerjee, A.~Mitra and P.~Mukherjee, \emph{A new formulation of non-relativistic diffeomorphism invariance}, \href{https://doi.org/10.1016/j.physletb.2014.09.047}{\emph{Phys. Lett. B} {\bfseries 737} (2014) 369} [\href{https://arxiv.org/abs/1404.4491}{{\ttfamily 1404.4491}}].

\bibitem{Banerjee:2015twa}
R.~Banerjee, A.~Mitra and P.~Mukherjee, \emph{General algorithm for nonrelativistic diffeomorphism invariance}, \href{https://doi.org/10.1103/PhysRevD.91.084021}{\emph{Phys. Rev. D} {\bfseries 91} (2015) 084021} [\href{https://arxiv.org/abs/1501.05468}{{\ttfamily 1501.05468}}].

\bibitem{Banerjee:2020wbr}
R.~Banerjee, \emph{Demystification of nonrelativistic theories in curved background}, \href{https://doi.org/10.1142/S021827182043015X}{\emph{Int. J. Mod. Phys. D} {\bfseries 29} (2020) 2043015} [\href{https://arxiv.org/abs/2008.05727}{{\ttfamily 2008.05727}}].

\bibitem{Banerjee:2020bym}
R.~Banerjee and P.~Mukherjee, \emph{Canonical formulation for a non-relativistic spinning particle coupled to gravity}, \href{https://doi.org/10.1088/1361-6382/abbf69}{\emph{Class. Quant. Grav.} {\bfseries 37} (2020) 235004} [\href{https://arxiv.org/abs/2003.05693}{{\ttfamily 2003.05693}}].

\bibitem{PhysRevD.111.026003}
A.~Fontanella and J.M.~Nieto~Garc\'{\i}a, \emph{Constructing nonrelativistic ${\mathrm{ads}}_{5}/{\mathrm{cft}}_{4}$ holography}, \href{https://doi.org/10.1103/PhysRevD.111.026003}{\emph{Phys. Rev. D} {\bfseries 111} (2025) 026003}.

\bibitem{Karan:2024jgn}
S.~Karan and B.R.~Majhi, \emph{{A time-like window into tensionless worldsheets}},  \href{https://arxiv.org/abs/2412.06387}{{\ttfamily 2412.06387}}.

\bibitem{Roychowdhury:2020nonrel_integrable}
D.~Roychowdhury, \emph{Nonrelativistic strings on and integrable systems}, \href{https://doi.org/10.1016/j.nuclphysb.2020.115220}{\emph{Nucl. Phys. B} {\bfseries 961} (2020) 115220} [\href{https://arxiv.org/abs/2003.02613}{{\ttfamily 2003.02613}}].

\bibitem{Roychowdhury:2020giant_magnons}
D.~Roychowdhury, \emph{Nonrelativistic giant magnons from newton cartan strings}, \href{https://doi.org/10.1007/JHEP02(2020)109}{\emph{JHEP} {\bfseries 02} (2020) 109} [\href{https://arxiv.org/abs/2001.01061}{{\ttfamily 2001.01061}}].

\bibitem{Andringa:2012uz}
R.~Andringa, E.~Bergshoeff, J.~Gomis and M.~de~Roo, \emph{{'Stringy' Newton-Cartan Gravity}}, \href{https://doi.org/10.1088/0264-9381/29/23/235020}{\emph{Class. Quant. Grav.} {\bfseries 29} (2012) 235020} [\href{https://arxiv.org/abs/1206.5176}{{\ttfamily 1206.5176}}].

\bibitem{Bergshoeff:2014uea}
E.~Bergshoeff, J.~Rosseel and T.~Zojer, \emph{Newton–cartan (super)gravity as a non-relativistic limit}, \href{https://doi.org/10.1088/0264-9381/32/20/205003}{\emph{Class. Quant. Grav.} {\bfseries 32} (2015) 205003} [\href{https://arxiv.org/abs/1505.02095}{{\ttfamily 1505.02095}}].

\bibitem{Bergshoeff:2019pij}
E.A.~Bergshoeff, J.~Gomis, J.~Rosseel, C.~Simsek and Z.~Yan, \emph{{String Theory and String Newton-Cartan Geometry}}, \href{https://doi.org/10.1088/1751-8121/ab56e9}{\emph{J. Phys. A} {\bfseries 53} (2020) 014001} [\href{https://arxiv.org/abs/1907.10668}{{\ttfamily 1907.10668}}].

\bibitem{Pal:2024btm}
S.K.~Pal and P.~Nandi, \emph{{Symmetry harmonization: exploring deformed oscillators and dissipative dynamics through the glass of Newton\textendash{}Hooke algebra}}, \href{https://doi.org/10.1140/epjc/s10052-024-12662-4}{\emph{Eur. Phys. J. C} {\bfseries 84} (2024) 312} [\href{https://arxiv.org/abs/2401.12957}{{\ttfamily 2401.12957}}].

\bibitem{PhysRevD.103.046020}
R.~Banerjee, S.~Moinuddin and P.~Mukherjee, \emph{New approach to the study of nonrelativistic bosonic string in flat spacetime}, \href{https://doi.org/10.1103/PhysRevD.103.046020}{\emph{Phys. Rev. D} {\bfseries 103} (2021) 046020}.

\bibitem{Moinuddin:2021oft}
S.~Moinuddin and P.~Mukherjee, \emph{{Geometry of nonrelativistic string}}, \href{https://doi.org/10.1088/1402-4896/acab3e}{\emph{Phys. Scripta} {\bfseries 98} (2023) 015304} [\href{https://arxiv.org/abs/2105.06218}{{\ttfamily 2105.06218}}].

\bibitem{Andringa:2010it}
R.~Andringa, E.~Bergshoeff, S.~Panda and M.~de~Roo, \emph{Newtonian gravity and the bargmann algebra}, \href{https://doi.org/10.1088/0264-9381/28/10/105011}{\emph{Class. Quant. Grav.} {\bfseries 28} (2011) 105011} [\href{https://arxiv.org/abs/1011.1145}{{\ttfamily 1011.1145}}].

\bibitem{Polchinski1998}
J.~Polchinski, \emph{String Theory, Volume I: An Introduction to the Bosonic String}, vol.~1 of \emph{Cambridge Monographs on Mathematical Physics}, Cambridge University Press, Cambridge (1998).

\bibitem{Banerjee:2018gqz}
R.~Banerjee and P.~Mukherjee, \emph{{Galilean gauge theory from Poincare gauge theory}}, \href{https://doi.org/10.1103/PhysRevD.98.124021}{\emph{Phys. Rev. D} {\bfseries 98} (2018) 124021} [\href{https://arxiv.org/abs/1810.03902}{{\ttfamily 1810.03902}}].

\bibitem{Banerjee:2017jyb}
R.~Banerjee, S.~Chakraborty, A.~Mitra and P.~Mukherjee, \emph{{Cosmological implications of shift symmetric Galileon field}}, \href{https://doi.org/10.1103/PhysRevD.96.064023}{\emph{Phys. Rev. D} {\bfseries 96} (2017) 064023} [\href{https://arxiv.org/abs/1705.06941}{{\ttfamily 1705.06941}}].

\bibitem{Bergshoeff:2018yvt}
E.~Bergshoeff, J.~Gomis and Z.~Yan, \emph{{Nonrelativistic String Theory and T-Duality}}, \href{https://doi.org/10.1007/JHEP11(2018)133}{\emph{JHEP} {\bfseries 11} (2018) 133} [\href{https://arxiv.org/abs/1806.06071}{{\ttfamily 1806.06071}}].

\bibitem{Gomis:2000bd}
J.~Gomis and H.~Ooguri, \emph{Nonrelativistic closed string theory}, \href{https://doi.org/10.1063/1.1372697}{\emph{Journal of Mathematical Physics} {\bfseries 42} (2001) 3127} [\href{https://arxiv.org/abs/hep-th/0009181}{{\ttfamily hep-th/0009181}}].

\bibitem{Dirac:1964}
P.A.M.~Dirac, \emph{Lectures on Quantum Mechanics}, Yeshiva University Press (1964).

\bibitem{Hanson:1976cn}
A.~Hanson, T.~Regge and C.~Teitelboim, \emph{Constrained hamiltonian systems}, {\emph{Accademia Nazionale dei Lincei, Rome} (1976) }.

\bibitem{Nandi:2018hww}
P.~Nandi, S.~Kumar~Pal and R.~Verma, \emph{{Particle dynamics and Lie-algebraic type of non-commutativity of space\textendash{}time}}, \href{https://doi.org/10.1016/j.nuclphysb.2018.08.008}{\emph{Nucl. Phys. B} {\bfseries 935} (2018) 183} [\href{https://arxiv.org/abs/1807.05062}{{\ttfamily 1807.05062}}].

\bibitem{Banerjee:1999yc}
R.~Banerjee, H.J.~Rothe and K.D.~Rothe, \emph{{Hamiltonian approach to Lagrangian gauge symmetries}}, \href{https://doi.org/10.1016/S0370-2693(99)00977-6}{\emph{Phys. Lett. B} {\bfseries 463} (1999) 248} [\href{https://arxiv.org/abs/hep-th/9906072}{{\ttfamily hep-th/9906072}}].

\bibitem{Banerjee:1999hu}
R.~Banerjee, H.J.~Rothe and K.D.~Rothe, \emph{{Master equation for Lagrangian gauge symmetries}}, \href{https://doi.org/10.1016/S0370-2693(00)00323-3}{\emph{Phys. Lett. B} {\bfseries 479} (2000) 429} [\href{https://arxiv.org/abs/hep-th/9907217}{{\ttfamily hep-th/9907217}}].

\bibitem{Maskawa:1976hw}
T.~Maskawa and H.~Nakajima, \emph{{Singular Lagrangian and Dirac-Faddeev Method: Existence Theorems of Constraints in Standard Forms}}, \href{https://doi.org/10.1143/PTP.56.1295}{\emph{Prog. Theor. Phys.} {\bfseries 56} (1976) 1295}.

\bibitem{PhysRevD.72.066015}
R.~Banerjee, P.~Mukherjee and A.~Saha, \emph{Bosonic $\mathrm{P}$-brane and arnowitt-deser-misner decomposition}, \href{https://doi.org/10.1103/PhysRevD.72.066015}{\emph{Phys. Rev. D} {\bfseries 72} (2005) 066015}.

\bibitem{PhysRevD.70.026006}
R.~Banerjee, P.~Mukherjee and A.~Saha, \emph{Interpolating action for strings and membranes: A study of symmetries in the constrained hamiltonian approach}, \href{https://doi.org/10.1103/PhysRevD.70.026006}{\emph{Phys. Rev. D} {\bfseries 70} (2004) 026006}.

\bibitem{Banerjee:2002ky}
R.~Banerjee, B.~Chakraborty and S.~Ghosh, \emph{{Noncommutativity in open string: A Gauge independent analysis}}, \href{https://doi.org/10.1016/S0370-2693(02)01944-5}{\emph{Phys. Lett. B} {\bfseries 537} (2002) 340} [\href{https://arxiv.org/abs/hep-th/0203199}{{\ttfamily hep-th/0203199}}].

\bibitem{Banerjee:2003tk}
R.~Banerjee, B.~Chakraborty and K.~Kumar, \emph{{Membrane and noncommutativity}}, \href{https://doi.org/10.1016/j.nuclphysb.2003.07.009}{\emph{Nucl. Phys. B} {\bfseries 668} (2003) 179} [\href{https://arxiv.org/abs/hep-th/0306122}{{\ttfamily hep-th/0306122}}].

\bibitem{Bagchi2013}
A.~Bagchi, \emph{Tensionless strings and galilean conformal algebra}, \href{https://doi.org/10.1007/JHEP05(2013)141}{\emph{Journal of High Energy Physics} {\bfseries 2013} (2013) 141} [\href{https://arxiv.org/abs/1303.0291}{{\ttfamily 1303.0291}}].

\bibitem{Bagchi2023}
A.~Bagchi, A.~Banerjee, J.~Hartong, E.~Have, K.S.~Kolekar and M.~Mandlik, \emph{Strings near black holes are carrollian}, \href{https://doi.org/10.1103/PhysRevD.110.086009}{\emph{Physical Review D} {\bfseries 110} (2024) 086009} [\href{https://arxiv.org/abs/2312.14240}{{\ttfamily 2312.14240}}].

\bibitem{Gomis:2019zyu}
J.~Gomis, J.~Oh and Z.~Yan, \emph{{Nonrelativistic String Theory in Background Fields}}, \href{https://doi.org/10.1007/JHEP10(2019)101}{\emph{JHEP} {\bfseries 10} (2019) 101} [\href{https://arxiv.org/abs/1905.07315}{{\ttfamily 1905.07315}}].

\bibitem{SW}
N.~Seiberg and E.~Witten, \emph{String theory and noncommutative geometry}, \href{https://doi.org/10.1088/1126-6708/1999/09/032}{\emph{Journal of High Energy Physics} {\bfseries 1999} (1999) 032} [\href{https://arxiv.org/abs/hep-th/9908142}{{\ttfamily hep-th/9908142}}].

\bibitem{PhysRevD.76.064007}
A.A.~Deriglazov, C.~Neves, W.~Oliveira, E.M.C.~Abreu, C.~Wotzasek and C.~Filgueiras, \emph{Open string with a background $b$ field as the first order mechanics, noncommutativity, and soldering formalism}, \href{https://doi.org/10.1103/PhysRevD.76.064007}{\emph{Phys. Rev. D} {\bfseries 76} (2007) 064007}.

\bibitem{LeBellac:1973unm}
M.~Le~Bellac and J.M.~L\'evy-Leblond, \emph{{Galilean electromagnetism}}, \href{https://doi.org/10.1007/BF02895715}{\emph{Nuovo Cim. B} {\bfseries 14} (1973) 217}.

\bibitem{Banerjee:2022eaj}
R.~Banerjee and S.~Bhattacharya, \emph{{New formulation of Galilean relativistic Maxwell theory}}, \href{https://doi.org/10.1103/PhysRevD.107.105022}{\emph{Phys. Rev. D} {\bfseries 107} (2023) 105022} [\href{https://arxiv.org/abs/2211.12023}{{\ttfamily 2211.12023}}].

\bibitem{PhysRevLett.133.151601}
A.~Fontanella and J.M.~Nieto~Garc\'{\i}a, \emph{Nonrelativistic holography from ${\mathrm{ads}}_{5}/{\mathrm{cft}}_{4}$}, \href{https://doi.org/10.1103/PhysRevLett.133.151601}{\emph{Phys. Rev. Lett.} {\bfseries 133} (2024) 151601}.

\bibitem{Blas:2010hb}
D.~Blas, O.~Pujolas and S.~Sibiryakov, \emph{{Models of non-relativistic quantum gravity: The Good, the bad and the healthy}}, \href{https://doi.org/10.1007/JHEP04(2011)018}{\emph{JHEP} {\bfseries 04} (2011) 018} [\href{https://arxiv.org/abs/1007.3503}{{\ttfamily 1007.3503}}].

\end{thebibliography}\endgroup

\end{document}